# A Nanomagnetic Voltage-Tunable Correlation Generator between Two Random Bit Streams for Stochastic Computing

M. T. McCray, Md Ahsanul Abeed, *Student Member, IEEE*, and Supriyo Bandyopadhyay, *Fellow, IEEE*

*Abstract*— **Graphical probabilistic circuit models of stochastic computing are more powerful than the predominant deep learning models, but also have more demanding requirements. For example, they require "programmable stochasticity", e.g. generating two random binary bit streams with *tunable* amount of correlation between the corresponding bits in the two streams. Electronic implementation of such a system would call for several components leaving a large footprint on a chip and dissipating excessive amount of energy. Here, we show an elegant implementation with just two dipole-coupled magneto-tunneling junctions (MTJ), with magnetostrictive soft layers, fabricated on a piezoelectric film. The resistance states of the two MTJs (high or low) encode the bits in the two streams. The first MTJ is driven to a random resistance state via a current or voltage generating spin transfer torque and/or voltage controlled magnetic anisotropy, while the second MTJ's resistance state is determined solely by dipole coupling with the first. The effect of dipole coupling can be varied with local strain applied to the second MTJ with a local voltage (~ 0.2 V) and that varies the correlation between the resistance states of the two MTJs and hence between the bits in the two streams (from 0% to 100%). This paradigm can be extended to arbitrary number of bit streams.**

## I. INTRODUCTION

Two or more random binary bit streams with *tunable* amount of correlation between them are needed for many models of stochastic computing, graphical circuit models, computer vision and image processing applications [1-4]. They can be produced with *software* (which is slow and consumes excessive processor and memory resources) or with complex *hardware* that involves shift registers, Boolean logic gates, random number generators, multiplexers, binary counters, comparators, etc. [4]. These hardware platforms consume large areas on a chip, dissipate enormous amounts of energy, and are expensive. In this paper, we show how two bit streams with *controlled amount of correlation* between corresponding bits in the two streams can be generated with two magneto-tunneling junctions fabricated on a piezoelectric film, whose soft layers are magnetostrictive and placed close enough to each other to be dipole-coupled. The correlation between the bits can be varied from 0% to 100% with a small voltage (~0.2 V) resulting in a low-power platform with a small system footprint.

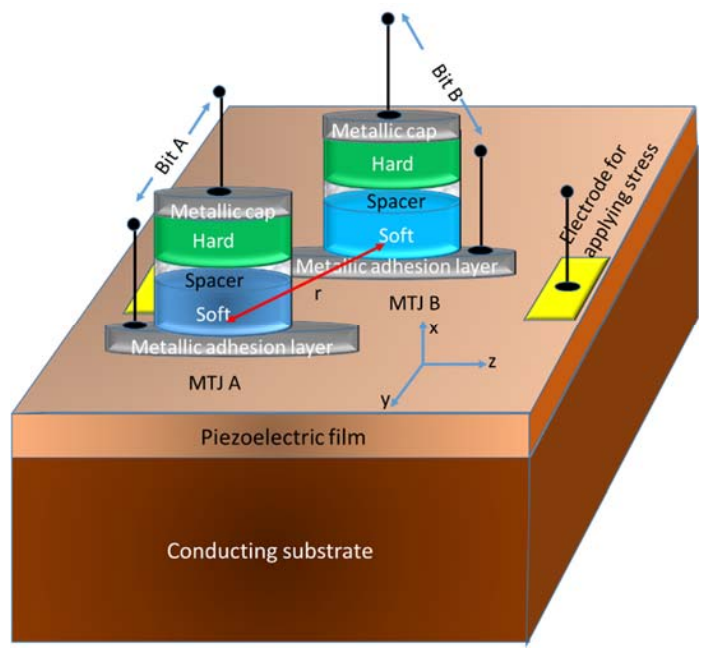

Fig. 1: A magneto-elastic system for generating two random bits with controlled amount of correlation.

Fig. 1 shows the system being discussed. It consists of two magneto-tunneling junctions (MTJs) of elliptical cross-section whose major and minor axes are mutually parallel. Both the hard and soft layers have in-plane magnetic anisotropy, but the principle described here would not be affected if they had perpendicular magnetic anisotropy. The MTJs are fabricated on a poled piezoelectric film deposited on a conducting substrate. Two electrodes are delineated on the piezoelectric film (flanking the soft layer of MTJ B) with appropriate dimensions and spacing to generate biaxial strain (tensile along the major

This work was supported in part by the U.S. National Science Foundation under grants ECCS-1609303 and CCF-1815033.

All authors are with the Department of Electrical and Computer Engineering, Virginia Commonwealth University, Richmond, VA 23284, USA

(emails: mccraymt@mymail.vcu.edu, abeedma@mymail.vcu.edu, sbandy@vcu.edu). The authors acknowledge fruitful discussions with Prof. Amit Ranjan Trivedi of the University of Illinois at Chicago.

axis and compressive along the minor axis, or vice versa) underneath the magnetostrictive soft layer of MTJ B when a voltage of appropriate polarity is applied between the electrodes and the bottom conducting substrate [5, 6]. By varying the voltage, we can vary the magnitude of the strain experienced by the magnetostrictive soft layer of MTJ B. The soft layer of MTJ A will also experience some strain from MTJ B, but its magnetization will be determined by the voltage applied across it, which will override any strain effect. No voltage is applied across MTJ B.

The spacing between the MTJs is made sufficiently small to have significant dipole coupling between the two soft layers.

The two binary bits (Bit A and Bit B) are encoded in the resistance states of the two MTJs – high resistance is bit 1 and low resistance bit 0.

## II. OPERATING PRINCIPLE

To generate a random Bit A, we apply a voltage across MTJ A which generates both voltage controlled magnetic anisotropy (VCMA) and also drives a spin polarized current through it, thereby generating spin transfer torque (STT). Depending on the magnitude of the applied voltage, we can generate different probability of switching the resistance of MTJ A as shown in ref. [7] (see Fig. 2 of this reference). For example, if we choose to make Bit A 0 with a certain probability (say, 70% probability of 0 and hence 30% of 1), we will first initialize the resistance state of Bit A to 0 with a large voltage of appropriate polarity and then apply another voltage of appropriate magnitude to make it flip with 30% probability, resulting in the desired probability distribution $P(A)$ [A = 0 or 1].

To generate the random Bit B that is *controllably correlated* with Bit A, we do not apply any voltage or current to MTJ B. Instead, we rely on the dipole coupling between the soft layers of MTJ A and MTJ B to set the resistance state of MTJ B, and hence the value of the Bit B, depending on the value of Bit A, thereby generating the conditional probability $P(B/A)$. One would expect that dipole coupling would always make the magnetization of the soft layer of MTJ B become antiparallel to the magnetization of the soft layer of MTJ A, resulting in perfect (anti-) correlation between bits A and B, but this does not happen in practice. There is a shape anisotropy energy barrier within the soft layer of MTJ B, which will *have to be overcome* by dipole coupling to make its magnetization rotate from its initial orientation (if it was not initially antiparallel) to become antiparallel to that of the soft layer of MTJ A. If the energy barrier cannot be overcome, then bit B will remain in its previous state and not be anti-correlated with bit A. We elucidate this in Fig. 2.

Suppose the magnetization of the soft layer of MTJ B is initially at $\theta = 0^0$ (see Fig. 2 for the definition of $\theta$) and the magnetization of MTJ A became oriented along $\theta = 0^0$ by the applied voltage. This places the two soft layers in the parallel configuration. The potential energy profile in the soft layer of MTJ B as a function of its magnetization orientation will look like curve I in the lower panel in Fig. 2. The system will be stuck in the in the metastable state (local energy minimum) at $\theta = 0^0$ (keeping MTJ A and MTJ B parallel) and not be able to transition to the global energy minimum at $\theta = 180^0$ (where MTJ A and MTJ B will be mutually anti-parallel) because of the intervening potential energy barrier. In this case, bit B will not be able to respond to bit A and the two bits will be *uncorrelated*. However, if we apply varying degree of stress to the soft layer of MTJ B, then we will depress the energy barrier by varying degree and thus be able to tune the probability that the system can transition to the ground state at $\theta = 180^0$ where the two MTJs will be anti-parallel and hence anti-correlated. Thus, by varying the voltage applied to the two electrodes in Fig. 1, which varies the stress produced in the soft layer of MTJ B, we can *vary the correlation between bits A and B* from no (0%) correlation to perfect (100%) anti-correlation.

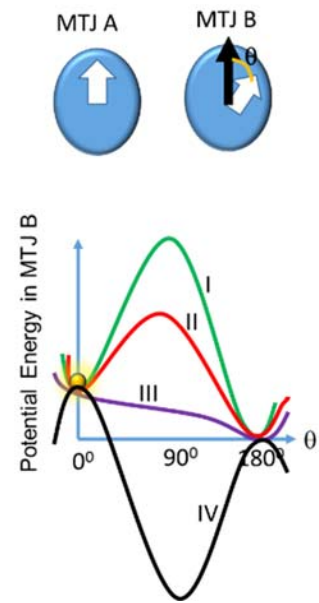

Fig. 2: Potential energy as a function of magnetization orientation $\theta$ within the soft layer of MTJ B in the presence of dipole coupling with MTJ A whose magnetization is oriented as shown in the top panel. The potential profile is asymmetric because of the dipole coupling. (I) No stress applied, (II) sub-critical stress applied, (III) critical stress applied, and (IV) super-critical stress applied. The ball indicates the initial magnetization state of the soft layer of MTJ B.

## III. MAXIMUM ANTI-CORRELATION AT CRITICAL STRESS

In Fig. 2, we see that different amounts of stress affect the energy barrier in different ways. We define "critical stress" as the amount of stress that *just* erodes the energy barrier completely, but does not invert it to create a potential well at $\theta = 90^0$ (Case III in Fig. 2). Sub-critical stress (Cases I and II) does not erode the potential barrier completely and super-critical stress (Case IV) inverts the potential barrier. Looking at the potential profiles, one would understand that if the initial magnetization state in MTJ B is parallel to the magnetization of MTJ A, which means that the initial state is at $\theta = 0^0$ as shown by the ball in Fig. 2, then the maximum likelihood that the magnetization of MTJ B will flip in response to the magnetization state of MTJ A (i.e. the ball will roll down smoothly to $\theta = 180^0$) will occur when the applied stress is the *critical stress.* Thus, *critical stress will result in the maximum anti-correlation.*

## IV. SIMULATION RESULTS

We have simulated the magneto-dynamics within the soft layers of both MTJ A and MTJ B in the presence of thermal noise by solving *coupled* stochastic Landau-Lifshitz-Gilbert (s-LLG) equations for the two soft layers. Both soft layers are made of Terfenol-D and have major axis = 100 nm, minor axis = 90 nm and thickness = 15 nm. The shape anisotropy energy barrier within each soft layer is 378 kT at room temperature.

In the case of MTJ A, the bit state is determined primarily by the spin polarized current that flows through the MTJ when the voltage is applied across it, and not by voltage controlled magnetic anisotropy (VCMA). Hence, we have ignored VCMA effects in our simulation and accounted only for the spin polarized current injection. The s-LLG equation is

$$\frac{d\vec{m}(t)}{dt} = -\gamma \vec{m}(t) \times \vec{H}_{total}(t) + \alpha \left( \vec{m}(t) \times \frac{d\vec{m}(t)}{dt} \right)$$
$$+ a\vec{m}(t) \times \left( \frac{\eta \vec{I}_s(t) \mu_B}{q M_s \Omega} \times \vec{m}(t) \right) + b \frac{\eta \vec{I}_s(t) \mu_B}{q M_s \Omega} \times \vec{m}(t) \qquad (1)$$

where

$$\hat{m}(t) = m_x(t)\hat{x} + m_y(t)\hat{y} + m_z(t)\hat{z} \qquad \left[ m_x^2(t) + m_y^2(t) + m_z^2(t) = 1 \right]$$

$$\vec{H}_{total} = \vec{H}_{demag} + \vec{H}_{stress} + \vec{H}_{dipole} + \vec{H}_{thermal}$$

$$\vec{H}_{demag} = -M_s N_{d-xx} m_x(t)\hat{x} - M_s N_{d-yy} m_y(t)\hat{y} - M_s N_{d-zz} m_z(t)\hat{z}$$

$$\vec{H}_{stress} = \frac{3}{\mu_0 M_s} \left( \lambda_s \sigma_{zz}(t) m_z(t) \right) \hat{z}$$

$$\vec{H}_{dipole} = \frac{M_s \Omega}{4\pi r^3} \left[ \tilde{m}_x \hat{x} - 2\tilde{m}_y \hat{y} + \tilde{m}_z \hat{z} \right] \qquad \begin{bmatrix} \text{tilda represents magnetization} \\ \text{of the neighboring nanomagnet} \end{bmatrix}$$

$$\vec{H}_{thermal} = \sqrt{\frac{2\alpha kT}{\gamma \left(1 + \alpha^2 \right) \mu_0 M_s \Omega (\Delta t)}} \left[ G_{(0,1)}^x(t)\hat{x} + G_{(0,1)}^y(t)\hat{y} + G_{(0,1)}^z(t)\hat{z} \right]$$

where $\alpha$ is the Gilbert damping factor of Terfenol-D ($\alpha = 0.1$), $r$ is the center-to-center separation between the soft layers, $G_{0,1}^i$ is a Gaussian of zero mean and unit standard deviation, $\gamma = 2\mu_B \mu_0 / \hbar$, $\mu_B$ is the Bohr magneton, $\mu_0$ is the permeability of free space, $M_s$ is the saturation magnetization of Terfenol-D ($8 \times 10^5$ A/m) [8], $\Omega$ is either soft layer's volume and $\Delta t$ is the time step used in the simulation (0.1 ps). We ignore magneto-crystalline anisotropy, assuming that the soft layers are amorphous.

Since it is difficult to incorporate biaxial strain in the s-LLG equation, we approximate it as a uniaxial stress directed along the major axis of the elliptical soft layer (z-axis; see Fig. 1). The uniaxial stress is written as $\sigma_{zz}$ and $\lambda_s$ is the saturation magnetostriction of Terfenol-D, which is 600 ppm [9]. The quantities $N_{d-ii}$ are the anisotropy coefficients which are found from the dimensions of the soft layer [10].

The voltage applied across MTJ A injects/extracts spin polarized carriers into its soft layer with their spins polarized in the $\pm z$ direction and $\eta$ is the spin polarization of the current, which we choose as 30%. In Equation (1), the last term on the right hand side is the field like torque and the second to last term is the Slonczewski torque. The coefficients $a$ and $b$ determine their relative weights and following ref. [11], we choose $a = 1$ and $b = 0.3$.

We define a *correlation parameter* as

$$C = \langle A \times B \rangle \qquad (2)$$

where $A$ is the value of bit A (either +1 or -1) and $B$ is the value of bit B (also either +1 or -1). We use -1 instead of 0 to represent the logic complement of the binary bit 1. The angular brackets denote ensemble average. The ensemble averaging is carried out over 1000 switching trajectories generated in our simulator. In each switching trajectory, Bit A assumes one value (+1 or -1) and B also assumes one value (+1 or -1). If the two bits are anti-correlated, their product is -1, whereas if they are uncorrelated, the product is +1. Thus, the ensemble average C (correlation parameter) is a *measure* of the degree of correlation between bits A and B. C= -1 indicates perfect (100%) anti-correlation and C = +1 indicates no (0%) correlation. Any intermediate value indicates partial correlation.

The switching trajectories are generated in the following manner. For MTJ A, we assume that the magnetization of its soft layer is initially pointing in the +z-direction. We then apply a voltage of appropriate polarity that results in passing a spin polarized current, with spin polarization in the –z-direction and magnitude varying between 30 and 40 mA (above the critical current for switching) and simulate the magneto-dynamics of the soft layer of MTJ A which yields its magnetization as a function of time $m_A(t)$ At the same time, we simulate the magneto-dynamics in the soft layer of MTJ B (which is initially magnetized in the –z-direction) to find its magnetization $m_B(t)$ as a function of time. The magneto-dynamics of the two magnets are coupled through the dipole coupling term; $m_A(t)$ determines $m_B(t)$ and vice versa. The simulations are carried out until the magnetizations of both soft layers reach steady state, at which point they are either approximately parallel or antiparallel. If they are parallel, then $A \times B = +1$, whereas if they are antiparallel, then $A \times B$ = -1. We average the value of $A \times B$ over the 1000 switching trajectories to calculate the correlation parameter defined in Equation (2).

In Fig. 3, we show plots of the correlation parameter as a function of the stress applied to the soft layer of MTJ B for different amplitudes of the current pulse injected into MJTJ A (no current is injected into MTJ B). The results are shown for four different center-to-center separations between the soft layers of MTJ A and MTJ B (and hence four different strengths of the dipole coupling; the dipole coupling strength varies as the inverse cube of the separation). There are several features of interest found in this figure. First and foremost, we find that we can vary the correlation parameter from +1 (no correlation or 0% correlation) to -1 (perfect anti-correlation or 100% correlation) continuously by varying the stress applied locally to the soft layer of MTJ B with the voltage impressed between the shorted electrodes in Fig. 1 and the conducting substrate. Thus, we can control the correlation between bits A and between 0% and 100% with an external voltage. We can do this for any pair of bits in the two bit streams and it is obvious that we can extend it to more than two bit streams by having additional dipole coupled MTJs. We can also have different spacings between the MTJs to achieve different correlations between different bit streams. This paradigm allows perfect tunability of the correlation between any two streams.

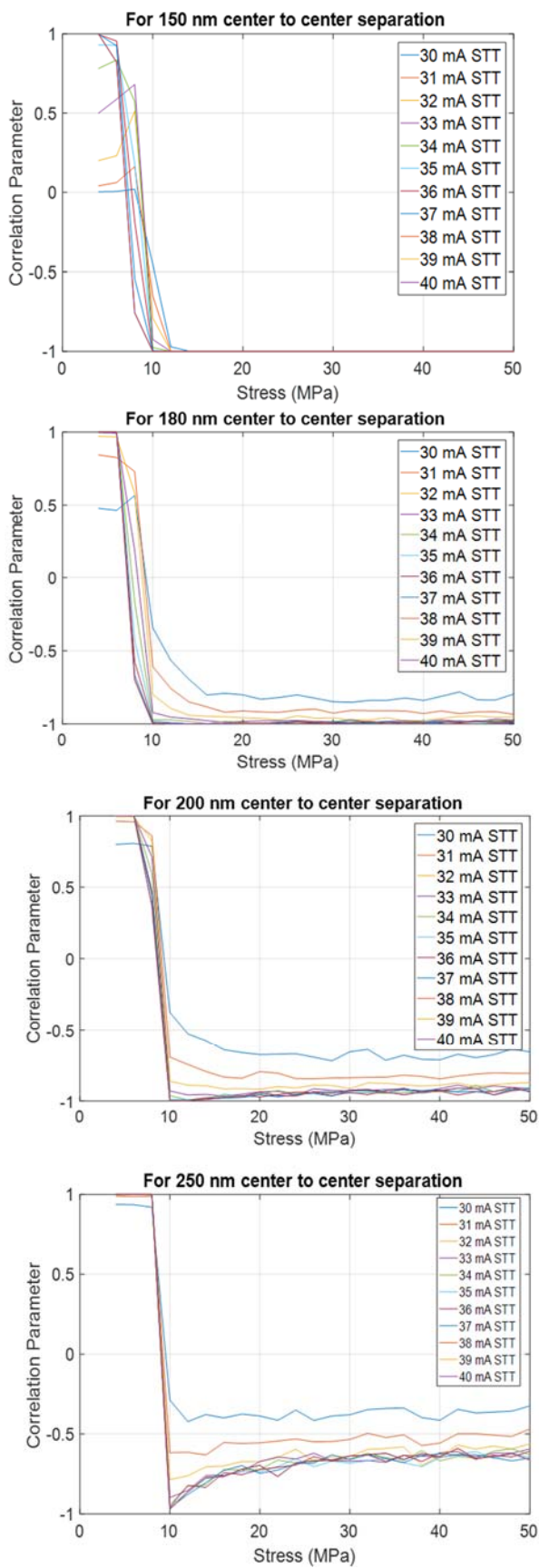

Fig. 3: Correlation parameter as a function of stress for different spin polarized currents and for four different spatial separations between MTJ A and MTJ B.

There are also other interesting features in Fig. 3. The anti-correlation decreases with increasing separation between the MTJs (i.e. decreasing dipole coupling strength), which is expected since the anti-correlation is caused by dipole coupling. The correlation parameter also depends on the current injected into MTJ A because this current determines the probability with which the state of Bit A is set. A low current may not succeed in "writing" the state of Bit A with high probability and a manifestation of that is the failure to reach perfect anti-correlation, even at high stress values, when the current is low. This feature becomes increasingly prominent with increasing separation between the MTJs (decreasing dipole coupling strength). The current densities used here are high ($\sim 5 \times 10^{12}$ A/m²), but this is a consequence of the materials and geometry chosen. Engineering the current densities to lower values is possible, but not an objective of this work.

Another very interesting feature clearly observed at higher inter-MTJ separations (bottom panels in Fig. 3) is that the anti-correlation becomes maximum at a certain stress value (note the non-monotonic behavior). This is the "critical stress" discussed in Section III and we see it clearly in these cases. Why the critical stress feature is more prominent at larger inter-MTJ separation (weaker dipole coupling) is also easy to understand by looking at Fig. 4. When the separation is larger (dipole coupling weaker), the energy difference between the local and global minima in the potential profile of the soft layer of MTJ B is smaller. In that case, it will be more important to completely erode the potential barrier without inverting it, in order to make the soft layer switch (ball to roll down from local to global minimum) with very high probability. On the other hand, if the inter-MTJ separation is smaller (dipole coupling stronger), then the energy difference between the local and global minima will be larger and eroding the barrier completely will not be so critical.

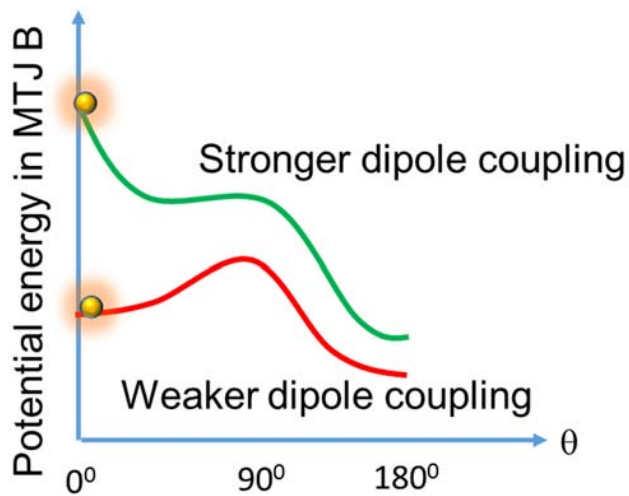

Fig. 4: Potential energy profile (energy as a function of magnetization orientation in the soft layer of MTJ B) for strong and weak dipole coupling.

Finally, we can roughly estimate how much voltage will be required to generate the stress of 10 MPa needed to vary the correlation from no correlation to nearly perfect anti-correlation (see Fig. 3). If we assume that the piezoelectric film is a 1 μm thick layer of PMN-PT, then the electric field $E$ needed to produce a uniaxial stress $\sigma_{zz}$ is given by

$$E = \frac{\sigma_{zz}}{d_{31}Y},\qquad (3)$$

where $d_{31}$ is the piezoelectric coefficient of PMN-PT and $Y$ is the Young's modulus of Terfenol-D. We will assume that all of the strain generated in the piezoelectric is transferred to the Terfenol-D soft layers. Using the values available in the literature for PMN-PT and Terfenol-D, $d_{31}$ = 1200 pm/V [12] and $Y$ = 45 GPa [9], the electric field needed to produce a stress of 10 MPa in the soft layers is 0.18 MV/m. This will be produced by a voltage of 0.18 V dropped over a film thickness of 1 μm. Thus, the modulating voltage needs to be no more than ~0.2 V.

The energy dissipated in changing the correlation from 0% to 100% is at most $CV^2$, where $C$ is the capacitance of the electrode pairs and $V$ is the voltage needed (0.18 V) to vary the correlation over the full range. The relative dielectric constant of a piezoelectric material like PMN-PT depends on the crystallographic orientation of the film, but is typically ~1,000 [13]. Assuming that the contact electrode areas are 1 μm × 1 μm and the piezoelectric film thickness is 1 μm, the capacitance $C$ ~ 8 fF. Therefore, the energy dissipated in tuning the correlation over its full range (0% to 100%) is a mere 260 aJ. That makes this an energy-efficient paradigm.

## V.  CONCLUSION

In this work, we have shown how to vary the correlation between corresponding bits in two random binary bit streams *controllably*, resulting in a *tunable probability correlator*. The correlation can be varied between 0% (no correlation) to 100% (perfect anti-correlation) by strain-engineering the effect of dipole coupling on the soft layer of the second MTJ with a voltage of only ~0.2 V, resulting in an energy dissipation of ~260 aJ. This is an energy-efficient paradigm.

The effect of dipole coupling could also have been varied with a local magnetic field applied selectively on the second MTJ, or by driving a controlled amount of spin polarized current through the second MTJ, but these approaches are much more energy-inefficient than applying strain. It is also very difficult to generate a localized magnetic field that is confined to the second MTJ and does not infringe on the first. That is why strain is the preferred modality.

A natural question to ask at this point is why the resistance state of the MTJ A is also not set by strain since it is more energy efficient than other modalities. If we apply super-critical stress to the soft layer of the MTJ A, then that will align its magnetization along its minor axis (see Fig. 2) and upon stress release, the magnetization will relax to either orientation along the major axis with equal probability. That will generate either a bit 1 or a bit 0 (or, equivalently, -1) with *equal* probability like a coin toss. There are two problems with this approach. First, we cannot set $P(A)$ to arbitrary value (e.g. 70% probability of bit 0 and 30% of bit 1); instead we are constrained to only 50:50 probability. Second, the strain fields under MTJ A and MTJ B may interfere (since they are placed so close together) and this could be a spoiler. That is why it is preferable to set A with a voltage instead of using strain.

Note that the paradigm works for MTJ spacing spanning a fairly large range from 150 nm to 250 nm. Hence, extreme lithographic precision is not required since there is a relatively large tolerance. MTJs with spacing in this range are not difficult to fabricate.

Bit streams with controlled amount of correlation between corresponding bits have applications in various areas of stochastic computing. While their traditional electronic implementation would require several components, here we have shown how they can be realized with just two dipole coupled magneto-tunneling junctions with magnetostrictive soft layers, fabricated on a piezoelectric film. This approach, which leverages coupled magneto-dynamics in two MTJs with strain-engineered dipole coupling effect in one of them, reduces device footprint, energy dissipation and cost.

**Mason T. McCray** is a high school senior from Maggie L. Walker Governor School for Government and International Studies at Richmond, VA who spent two semesters interning in the Department of Electrical and Computer Engineering at Virginia Commonwealth University where he carried out this work.

**Md Ahsanul Abeed** is a Ph.D. student in the Department of Electrical and Computer Engineering working on straintronic devices and systems. He received his Bachelor's degree from Bangladesh University of Engineering and Technology.

**Supriyo Bandyopadhyay** (SM'80, M'86, SM' 88, F' 05) is Commonwealth Professor at Virginia Commonwealth University. One of his research interests is straintronic devices and systems.